\documentclass[twocolumn,showpacs,amsmath,prb,superscriptaddress]{revtex4-1}
\usepackage{color}
\usepackage{graphicx}
\usepackage{epsfig,psfrag}
\usepackage{amsmath}
\usepackage{amssymb}
\usepackage{amsbsy}
\usepackage{amsthm}
\usepackage{amsfonts}
\usepackage{wasysym}
\usepackage{bbm}
\usepackage{tabularx}
\usepackage{euscript}
\usepackage{enumerate}
\usepackage{amsfonts}
\usepackage{exscale}
\usepackage{bbold}
\usepackage{float}
\usepackage{slashed}
\usepackage{subfigure}
\usepackage{comment}
\usepackage[colorlinks,citecolor=blue]{hyperref}

\usepackage[utf8]{inputenc}
\usepackage[T1]{fontenc}
\inputencoding{latin1}
\inputencoding{utf8}

\newcommand{\bea}{\begin{eqnarray}}
\newcommand{\eea}{\end{eqnarray}}
\newcommand{\ba}{\begin{array}}
\newcommand{\ea}{\end{array}}

\newcommand{\nn}{\nonumber\\}

\renewcommand{\Im}{{\rm Im}}
\renewcommand{\Re}{{\rm Re}}

\def\bea{\begin{eqnarray}}
\def\eea{\end{eqnarray}}

\def\nn{\nonumber}

\begin{document}

\title{Chiral topological superconductivity in Josephson junction}

\author{Shao-Kai Jian}
\affiliation{Condensed Matter Theory Center, Department of Physics, University of Maryland, College Park, Maryland 20742, USA}

\author{Shuai Yin}
\affiliation{School of Physics, Sun Yat-Sen University, Guangzhou 510275, China}

\begin{abstract}
We consider a heterostructure of semiconductor layers sandwiched between two superconductors, forming a two-dimensional Josephson junction. 
Applying a Zeeman field perpendicular to the junction can render a topological superconducting phase with the chiral Majorana edge mode. 
We show that the phase difference between two superconductors can efficiently reduce the magnetic field required to achieve the chiral topological superconductivity, providing an experimentally feasible setup to realize chiral Majorana edge modes. 
We also construct a lattice Hamiltonian of the setup to demonstrate the chiral Majorana edge mode and the Majorana bound state localized in vortices.
\end{abstract}
\date{\today}
\maketitle

\section{Introduction}

Since the discovery of topological insulators, intensive theoretical and experimental studies on various kinds of symmetry protected topological order have arisen~\cite{Kitaev2009, Ludwig2010}, among which the pursuit of Majorana fermions and Majorana zero modes is one of the most interesting and important issues. 
Owing to its non-Ablian statistics, the Majorana zero mode is of great potential to implement topological qubits in fault-tolerant topological quantum computations~\cite{Kitaev2003, Alicea2011, Halperin2012, DasSarma2015, Franz2015}. 
The Majorana zero mode can be realized in 5/2 fractional quantum Hall states~\cite{Stern2010, Read2000}, vortices of spinless $p+ip$ superconductors~\cite{Kitaev2006} and superconductor-semiconductor heterostructures~\cite{Fu2008, Sau2010, Alicea2010, DasSarma2010} in two dimensions, the edge of a ferromagnetic atomic chain on a superconductor~\cite{Kitaev2001, Yazdani2013, Simon2013}, and the planar Josephson junction~\cite{Stern2017, Marcus2019, Yacoby2019, Mayer2019}, etc., leading to ongoing interest among both condensed matter physics and quantum computation communities.

On the other hand, one-dimensional chiral Majorana fermion---the simplest solution satisfying the real version of the Dirac equation originally proposed by Ettore Majorana in 1937~\cite{Majorana1937}---also attracts lots of attentions in recent years. 
Realization of the chiral Majorana fermions in real world is not only of great theoretical importance, but more significantly also opens a new avenue for topological quantum computation~\cite{Lian2018}. 
The chiral Majorana fermion can be realized at the edge of the two-dimensional chiral topological superconductor in class D~\cite{Ludwig2010}. 
The chiral topological superconductor is a real analog of an integer quantum Hall insulator, where the number of chiral Majorana edge modes reflects the BdG Chern number of the occupied bands. 
A series of theoretical proposals arises including the superconductor-semiconductor heterostructure~\cite{Sau2010}, and superconductor-quantum anomalous Hall insulator heterostructure~\cite{Qi2010}, and so on. 
Although various experimental works have tried to realize the later proposal~\cite{He2017}, debates remain about whether true chiral Majorana edge modes are observed~\cite{Wen2018, Sau2018, Chang2019, Wang2019}.

\begin{figure}
	\centering
\subfigure[]{\label{setup}
	\includegraphics[width=0.24\textwidth]{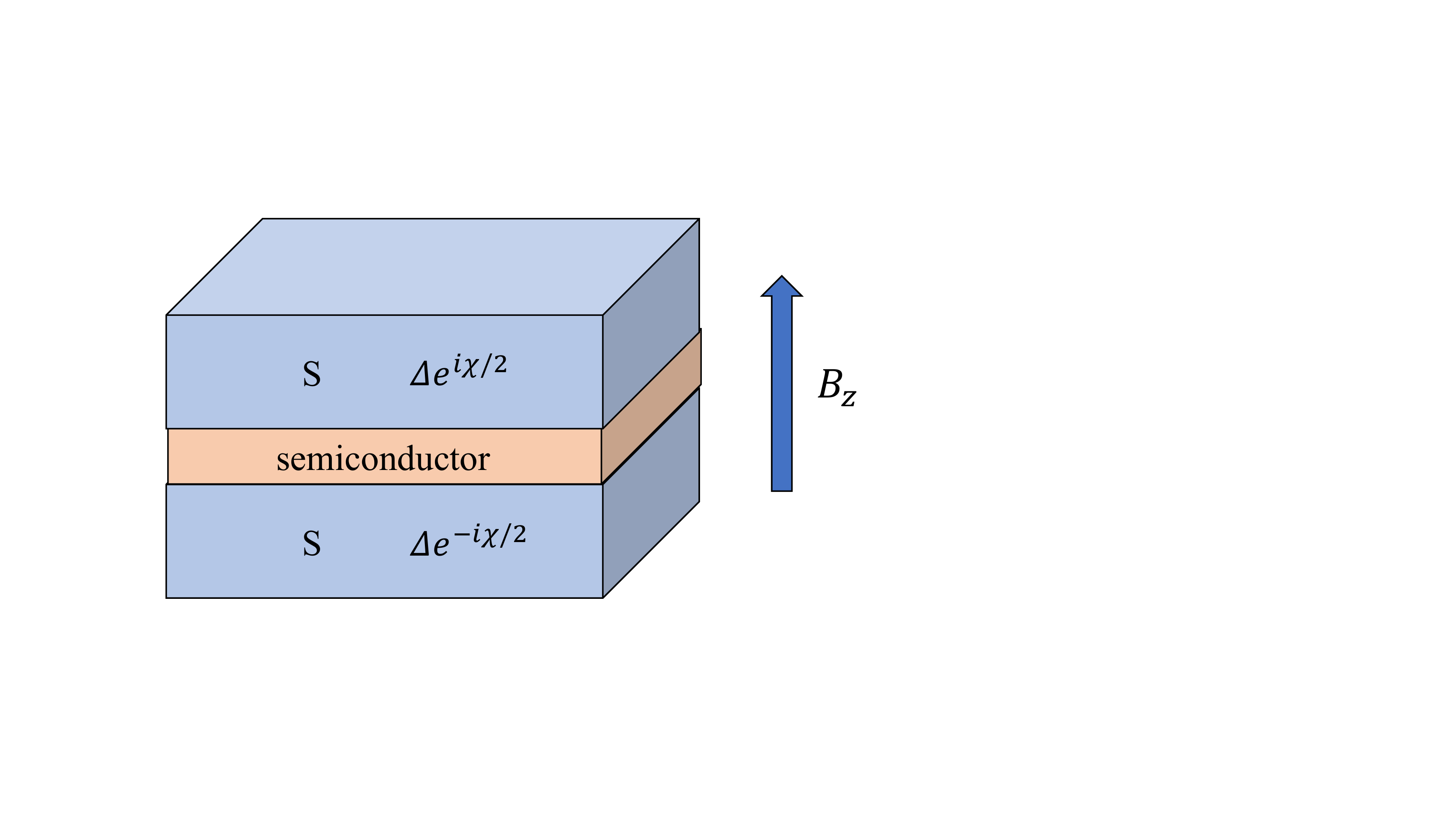}} 
\subfigure[]{\label{junction}
	\includegraphics[width=0.22\textwidth]{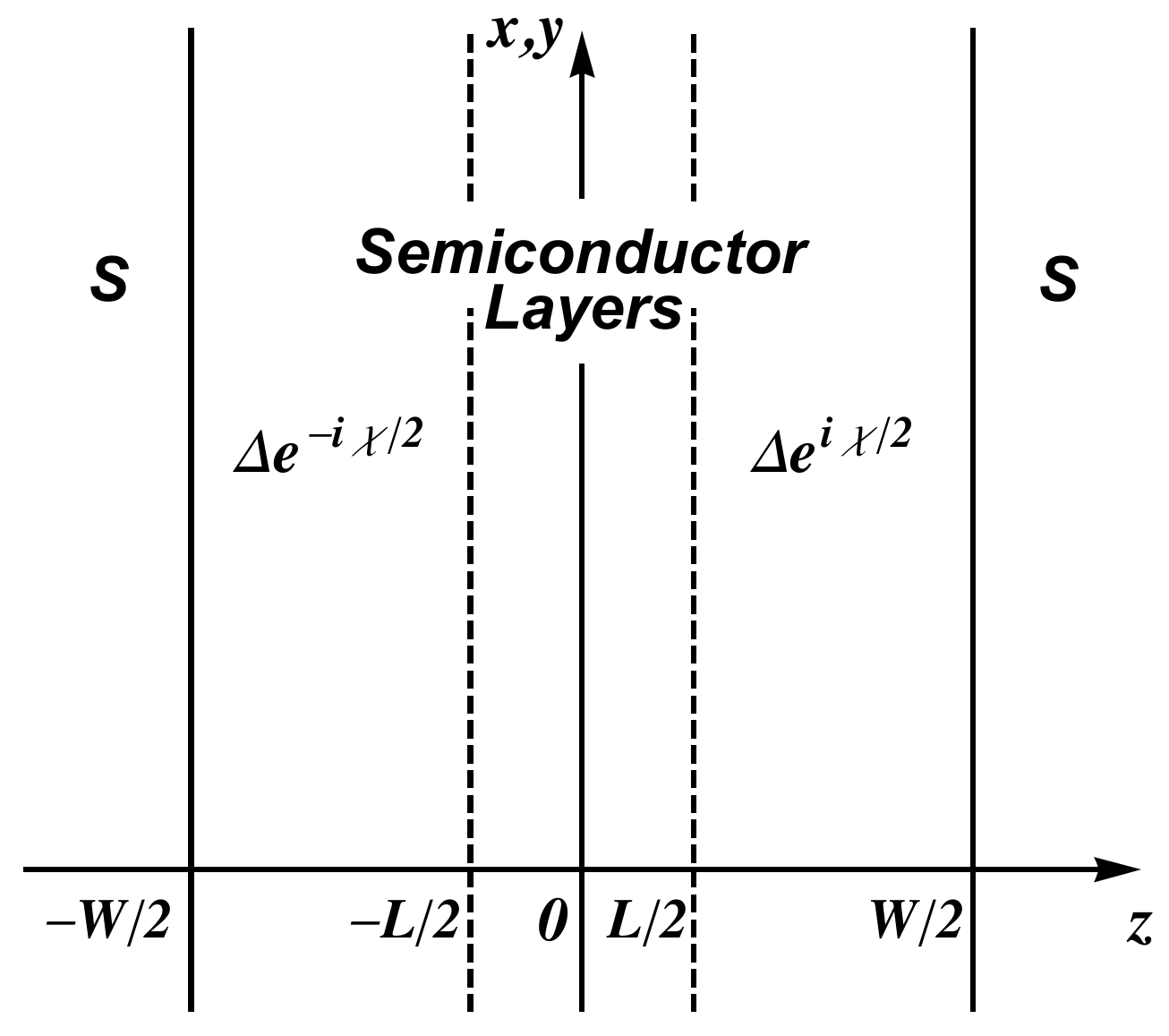}}
	\caption{\label{josephson}(a) A schematic plot of the setup. A few layers of semiconductor thin film is sandwiched between two superconductors (S). The heterostructure form a Josephson junction. The order parameters of two superconductors are given by $\Delta e^{\pm i\chi/2}$.  A out-of-plane magnetic field denoted by $B_z$ is also applied to tune the topological superconductivity. As we discussed in the main text, the Zeeman energy can be obtained by using a ferromagnetic semiconductor without applying an external magnetic field. (b) The proximitized SC state forms a domain wall in the semiconductor layers. The width of the domain wall and of the semiconductor layers are denoted by $L$ and $W$, respectively.}
\end{figure}

In this paper, we propose a different strategy to realize the chiral Majorana edge mode by taking advantage of the phase difference in the Josephson junction. 
As shown in Fig.~\ref{setup}, we consider a few layers of semiconductor sandwiched between two superconductors, forming a two-dimensional Josephson junction with a superconducting phase difference $\chi$. 
In the presence of a Zeeman field perpendicular to the junction, the phase difference $\chi$ provides a useful knob to tune the topological transition from a trivial superconductor to a chiral topological superconductor with a Majorana edge mode. 
In stead of the external magnetic field, the Zeeman energy can also be obtained by using a ferromagnetic semiconductor in our setup.
Namely, we can consider the superconductor/ferromagnetic semiconductor/superconductor heterostructure without applying the external magnetic field.
The advantage of this setup is that (i) the critical Zeeman field to achieve the topological transition is efficiently reduced by the phase difference $\chi$ without destroying the superconducting order and (ii) the transition can be realized without carefully gating the system. 
The continuous topological transition is described by two-dimensional Majorana cone at $\Gamma$ point, the center point of Brillouin zone. 
We also construct a lattice model to show the chiral Majorana edge mode and the Majorana zero bound states within the vortices---the two essential signatures of chiral topological superconductors.

The paper is organized as follows. 
In Section~\ref{sec:setup} we describe the Josephson junction setup and obtain the phase diagram using the scattering theory. 
A remarkable result is the critical Zeeman energy near the phase difference $\pi$ is in the order of $\Delta^2/\mu$, where $\mu$ and $\Delta$ are the chemical potential and the superconducting gap respectively, which is a significant improvement compared to previous proposals~\cite{Sau2010}.
In Section~\ref{sec:transition}, we reveal how the topological phase transition to the topological superconducting phase is achieved by showing that the gap closing point is described by a real version of the Dirac Hamiltonian that changes BdG Chern number.
The calculation illuminates the robustness of our mechanism to realize the topological transition.
In Section~\ref{sec:gap}, we show the gap of the topological superconducting phase is sizable, i.e., is in the same order of the proximitized superconducting gap, over nearly the full phase diagram.
In Section~\ref{sec:lattice}, a microscopic lattice model of our setup is constructed to demonstrate the phase diagram, the nontrivial chiral Majorana edge states and the Majorana zero modes in the topological phase.
Finally, we summarize the results in various aspects and also make a brief comparison with the previous heterostructure setup to show the advantage of our proposal in Section~\ref{sec:conclusion}.

\section{Josephson junction setup and the phase diagram} \label{sec:setup}

We consider a two-dimensional Josephson junction shown in Fig.~\ref{setup}, where a few layers of semiconductor are sandwiched between two superconductors with a superconducting phase difference $\chi$. 
Due to the proximity effect, the superconducting orders will penetrate into the semiconductor layers and form a domain wall in the semiconductor layers. 
One may wonder what the domain wall means since the superconducting order parameter is not a discrete variable.
More precisely, we are considering the situation that the width of the semiconductor is larger than twice of the superconducting proximity length, such that in the middle of the semiconductor, the proximitized superconducting gap is zero and the semiconductor is in the normal state.
We refer to the region with normal states the domain wall.
As shown in Fig.~\ref{junction}, the width of the domain wall and the semiconductor layers are $L$ and $W$, respectively. 
The superconducting order in the semiconductor layers is modeled by
\bea\label{SC}
	\Delta(z) =  \Delta e^{-i\chi/2} \Theta(-z-L/2) + \Delta e^{i\chi/2} \Theta(z-L/2),
\eea
where $\Delta$ ($\chi$) is the magnitude (the phase difference) of the superconductivity, and $\Theta$ is the step function, i.e., $\Theta(x) = 0 $ for $x<0$ and $\Theta(x) = 1 $ for $x \ge 0$. 
$z$ denotes the direction perpendicular to the junction. 
We assume that inside the domain wall $|z| \le L/2$ the superconducting order is zero. 
Because the proximity length of superconductivity is in general quite large, and the proximitized superconducting order in semiconductors opens up an energy gap, the amplitude of wavefunctions inside the semiconductor will concentrate near the domain wall, one can focus on the superconducting-normal state interface at $\pm L/2$ and take $W/L \rightarrow \infty$.

The BdG Hamiltonian of the semiconductor layers is $H = \Psi^\dag H_{BdG} \Psi$, where $\Psi= (\psi_\uparrow, \psi_\downarrow, \psi^\dag_\downarrow, - \psi^\dag_\uparrow)$, and
\bea
	H_{BdG}({\bf k}, z) =& \Big(\frac{{\bf k}^2 - \partial_z^2}{2m} -\mu\Big) \tau^z + H_{SOC}({\bf k}) \tau^z \nn \\
		& + (\Delta(z) \tau^+ + H.c.) + E_Z \sigma^z ,
\label{HBdG}
\eea
where ${\bf k} = (k_x, k_y)$ denotes the momenta in the junction plane, $\tau$ ($\sigma$) is the Pauli matrix acting on the Nambu (spin) space and $\tau^\pm \equiv \frac12(\tau^x \pm i \tau^y)$. 
$E_Z$ denotes the Zeeman energy induced by the external magnetic field perpendicular to the junction (along $z$ direction) or by using ferromagnetic semiconductor with out-of-plane magnetic orders. $H_{SOC}$ is the Dresselhaus spin-orbit coupling which will be discussed later.

The phase transition between the trivial and topological superconductors is characterized by a gap closing at $\bf{k}=0$ and consequently a change of the BdG Chern number. 
So, to look for the topological transition, we calculate the junction spectrum in the $\bf{k}=0$ subspace by the scattering theory~\cite{Beenakker1991}. 
As the spin is conserved during the scattering, we can focus on $\sigma^z=1$ subspace.
The transmission amplitude through the domain wall $-L/2<z<L/2$ is  
\bea 
	T = \left( \ba{cccc} e^{i k^e L} & 0 \\ 0 & e^{-i k^h L} \ea \right),
\eea 
where $k^{e/h}= \sqrt{2m(\mu \pm (\epsilon -  E_Z))}$ is the wave-vector of the normal state wavefunction and $\epsilon$ is the eigen-energy. 
By matching the wavefunction at the superconducting-normal state interface at $z= \pm L/2$ (see Appendix~\ref{append:scattering} for more details), the scattering matrix is obtained to be $S_{\pm L/2}= e^{\pm i \chi/4 \tau^z} S e^{\mp i \chi/4 \tau^z}$, where
\bea \label{eq:scattering_matrix}
	S = e^{-i\phi_S} \left( \ba{cccc} r e^{i\phi_N} & \sqrt{1-r^2} \\
	\sqrt{1-r^2} & -r e^{-i\phi_N} \ea \right),
\eea
where $r$ is the normal reflection amplitude, and $\phi_{S, N}$ are two phases given in the Appendix~\ref{append:scattering}. 
Using the scattering and transmission matrix, the spectrum of the bound state is determined by~\cite{Beenakker1991} $\det(1- S_{-L/2}TS_{+L/2}T)=0$, i.e.,
\bea 
	&&\cos[(k^e-k^h)L-2\phi_S] \nn \\
	&& \quad\quad\quad = (1-r^2) \cos \chi + r^2 \cos[(k^e + k^h)L + 2\phi_N].
\eea

Assuming a weak superconducting pairing $E_Z \lesssim \Delta \ll \mu$, the zero mode is given by
\bea\label{boundary}
	\arccos \frac{E_Z}{\Delta}= \frac{\tilde{\chi}}2 + \frac{\pi}2 \frac{E_Z}{E_T} + n \pi,
\eea
where $n\in \mathbb{Z}$, $\tilde \chi $ is a function of $\chi$,
\bea\label{chi}
	\tilde\chi(\chi) = \arccos \Big[(1- r^2) \cos \chi + r^2 \cos 2 k_F L \Big],
\eea
and $E_T \equiv (\pi/2) v_F/L$ is the Thouless energy of the junction. 
Here $k_F = \sqrt{2m \mu}$ and $v_F = \sqrt{2\mu/m}$. 
Note the domain wall width is in the order of the lattice constant, thus, the Thouless energy is in the order of the chemical potential, $E_T \sim \mu$, which is the largest energy scale. 

\begin{figure}
	\includegraphics[width=5cm]{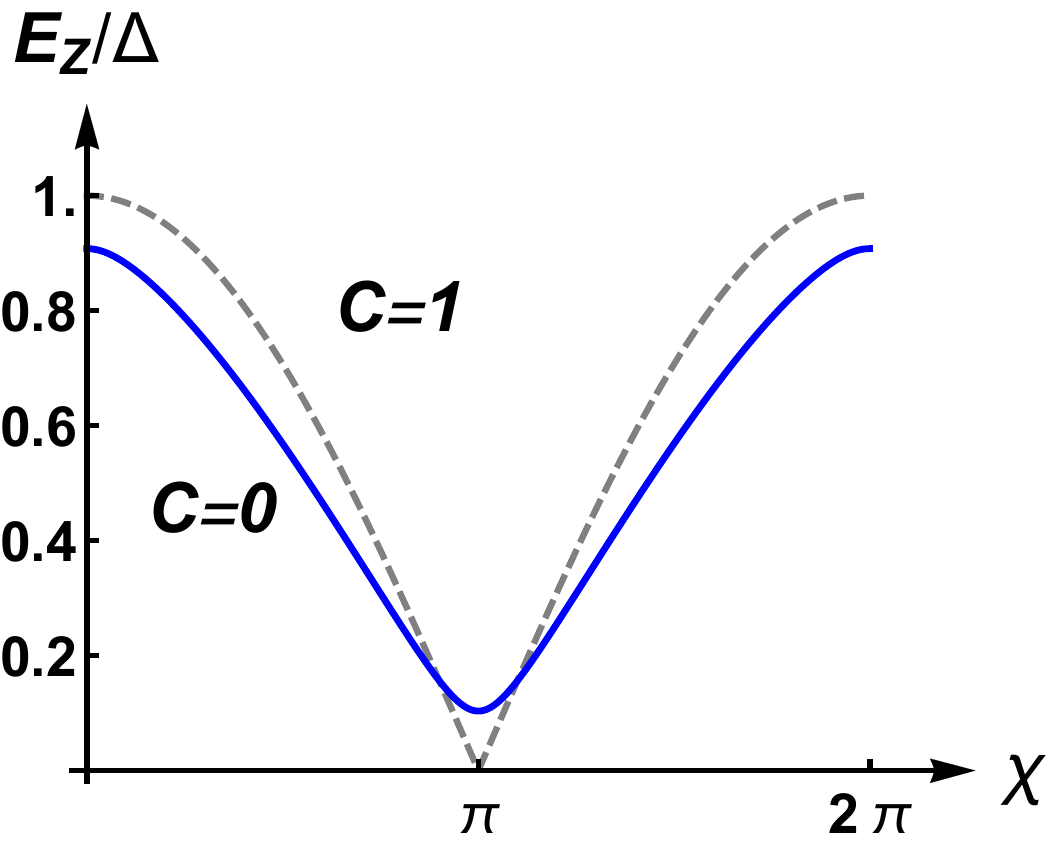}
	\caption{\label{phaseDiagram1}The phase diagram of the Josephson junction as a function of the junction phase difference and the Zeeman energy. $C$ denotes the BdG Chern number, where $C=0$ ($C=1$) implies the trivial (topological) superconductivity. The dashed (solid) line is the phase boundary without (with) normal reflection.}
\end{figure}

In the weak superconducting pairing limit, $\Delta \ll \mu$, the normal reflection~\cite{Beenakker1991} and the effect from Thouless energy, i.e. the second term on the right-hand side of Eq.~(\ref{boundary}), can be neglected. 
Then phase diagram is simply given by $\frac{E_Z}{\Delta} = |\cos \frac{\chi}2|$, as shown by the dashed line in Fig.~\ref{phaseDiagram1}. 
It is remarkable that at $\chi=\pi$, the critical Zeeman energy is reduced to zero, namely, an infinitesimal Zeeman field can turn the system into the chiral topological superconducting phase. 
Including corrections from the normal reflection to the lowest nontrivial order of $\Delta/\mu$, $r= \Delta/(2\mu) + \mathcal{O}(\Delta^2/\mu^2)$, the phase diagram is modified to the solid line in Fig.~\ref{phaseDiagram1}. 
Especially, at $\chi=\pi$, the correction is given by $\tilde{\chi}(\chi=\pi) = \pi - \Delta/\mu \cos k_F L$, so the critical Zeeman energy at $\chi = \pi$ is modified to be $\Delta^2/\mu$, which is still small at weak pairing limit, making it an experimentally achievable way to realize the chiral topological superconductivity.

\section{Topological phase transition} \label{sec:transition}

It is well known that spin-orbit coupling is crucial in the realization of topological phases. 
Direct semiconductors with an inversion-asymmetric zinc blende structure (point group $T_d$), such as GaAs, InSb, CdTe etc, have considerable size of spin-orbit coupling and are common in making quantum structures~\cite{WinklerBook}. 
Moreover, many of these materials have a similar band structure with the smallest gap at the $\Gamma$ point. 
To give a concrete example of spin-orbit couplings, we consider a heterostructure consisting of a few layers of such semiconductors grown in (001) direction. 
Because of the phase difference in the Josephson junction, the heterostructure breaks inversion symmetry. 
However, such a structure inversion asymmetry cannot induce Rashba spin-orbit coupling, since the system respects the composite $S_4$ and time reversal symmetry. 
As a result, the symmetry allowed spin-orbit coupling is the Dresselhaus term~\cite{WinklerBook, Dresselhaus1954}, $H_{SOC}({\bf k}) = \alpha (k_x \sigma^x - k_y \sigma^y)$, where $\alpha$ is the strength of spin-orbit coupling.

In the following, we will show that the topological phase transition is described by a two-dimensional Majorana cone---a real version of the Dirac fermion. 
To simplify the question, we set the width of domain wall $L$ to be zero. 
At the $\bf k =0 $ subspace of the normal semiconductor without the proximitized superconducting order and the external magnetic field, a continuous kinetic term of the one Kramers pair near the Fermi points, $\psi_{1,\uparrow}$ and $\psi_{2,\downarrow}$, is described by one-dimensional Dirac Hamiltonian $ -i v_F \partial_z \sigma^z$ (see Appendix~\ref{append:transition} for more details). 
Adding the proximitized SC order and the Zeeman field, the BdG Hamiltonian in Nambu basis $\Psi = (\psi_{1,\uparrow}, \psi_{2,\downarrow}, \psi_{1,\uparrow}^\dag, \psi_{2,\downarrow}^\dag)^T$ is given by (note the basis difference in Eq.~(\ref{HBdG})), $\mathcal{H}_{BdG}({\bf k}, z) =\mathcal H_0(z) +\mathcal H_1({\bf k})$, where
\bea
	&&\mathcal H_0(z) = -i v_F \partial_z \sigma^z - \Delta' \sigma^y \tau^y - \Delta''(z) \sigma^y \tau^x  + E_Z \sigma^z \tau^z, \label{H0}\nn\\ \\
	&&\mathcal H_1({\bf k}) = \frac{k_x^2 + k_y^2}{2m} \tau^z + \alpha(k_x \sigma^x - k_y \sigma^y \tau^z), \label{H1}
\eea
and $\Delta'= \Delta \cos(\chi/2)$ and $\Delta''(z) = \Delta \sin (\chi/2) \text{sgn}(z)$ refer to the real and imaginary part of the superconducting order, respectively. 
We have separated the Hamiltonian into an unperturbed part $\mathcal H_0$ and a perturbation $\mathcal H_1$, respectively.

In Eq.~(\ref{H0}), because of the domain wall formed in $\Delta''(z)$~\cite{Jackiw1976}, we get two bound states with eigen-energy $ E_{\pm}= \pm (E_Z - \Delta \cos \chi/2)$ that are related by particle-hole transformation,
\bea
	\psi_+(z) = f(z)  (1, 0, 0, -1)^T,  \quad \psi_-(z) = f(z)  (0, 1, -1, 0)^T, \nn\\
\eea
where $f(z) = (\mathcal N)^{-1/2}\exp[-\int_0^z dz' \frac{\Delta''(z')}{v_F}]$, and $\mathcal N$ is the normalization factor.

Now we consider the perturbation Eq.~(\ref{H1}) to the bound state $\psi_\pm$. 
Using the first-order perturbation theory, it is straightforward to get the effective Hamiltonian in the bound states subspace $(\psi_+, \psi_-)^T$,
\bea \label{eq:dirac}
	\mathcal H_\text{eff} = \left(  \ba{cccc} \delta m & -\alpha (k_x+ i k_y) \\
	-\alpha (k_x - i k_y) & -\delta m \ea \right),
\eea
where the mass is $\delta m = E_Z - \Delta \cos \chi/2 $. 
At critical point $\delta m=0$, the Hamiltonian describes a massless Majorana cone, corresponding to the phase transition from a trivial to a topological superconductor.

Similar analysis can be applied to the other Kramers pair $\psi_{1,\downarrow}$ and $\psi_{2,\uparrow}$, which shows that the critical point occurs at $E_Z/\Delta = -\cos \chi/2$. Combined with the above results, the phase boundary is $E_Z/\Delta = |\cos \chi/2|$, (we assume $E_Z>0$), consistent with the results from the scattering theory. 

The way that the topological transition is realized in Eq.~(\ref{eq:dirac}) is very robust in the following sense.
So long as there exists some mechanism to achieve a gap closing at the $\Gamma$ point, the symmetry allowed spin-orbit coupling is sufficient to render a nontrivial topological transition because the gap closing point is effectively described by a Dirac Hamiltonian.
The tunability of the domain wall bound state in $\bf k = 0$ subspace provides a useful mechanism to realize the gap closing.
We also want to emphasize the fact that the symmetry in our setup only allows that the Dresselhaus spin-orbit coupling is not essential.
The presence of other kinds of spin-orbit couplings, such as the Rashba spin-orbit coupling, will not affect the topological phase transition in general.

\begin{figure*}[htbp]
	\centering
	\subfigure[]{\label{phaseDiagram3}
		\includegraphics[height=3cm]{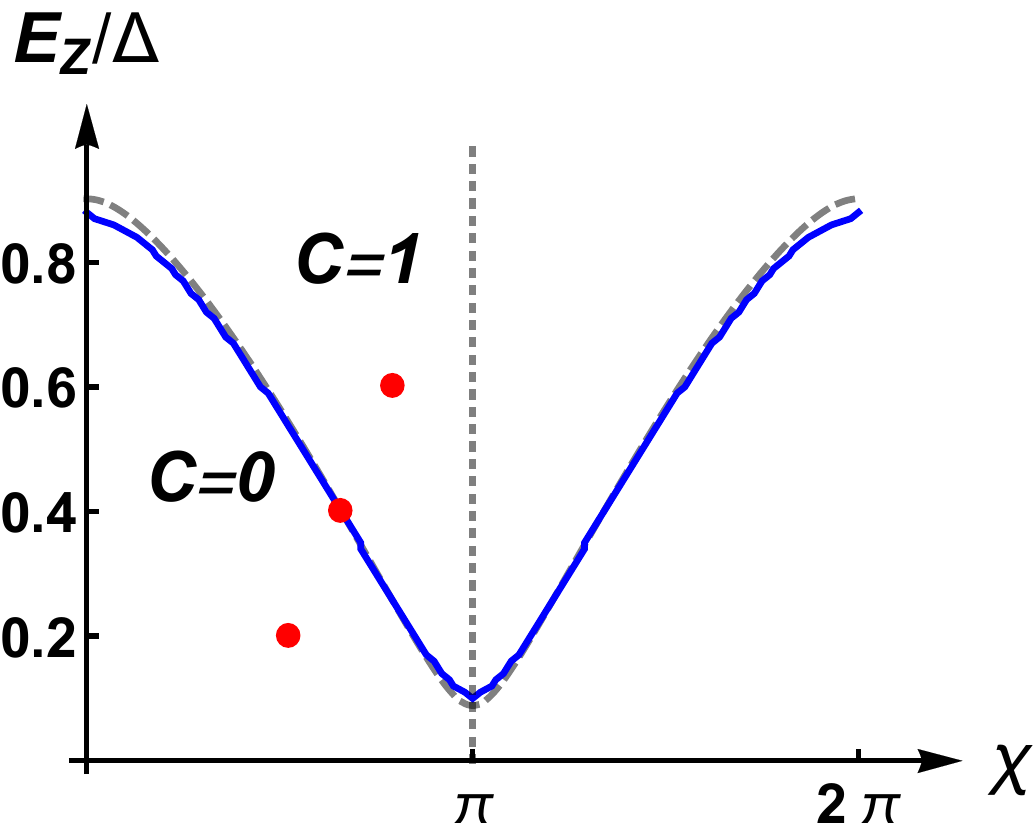}} \quad
	\subfigure[]{
		\includegraphics[height=3cm]{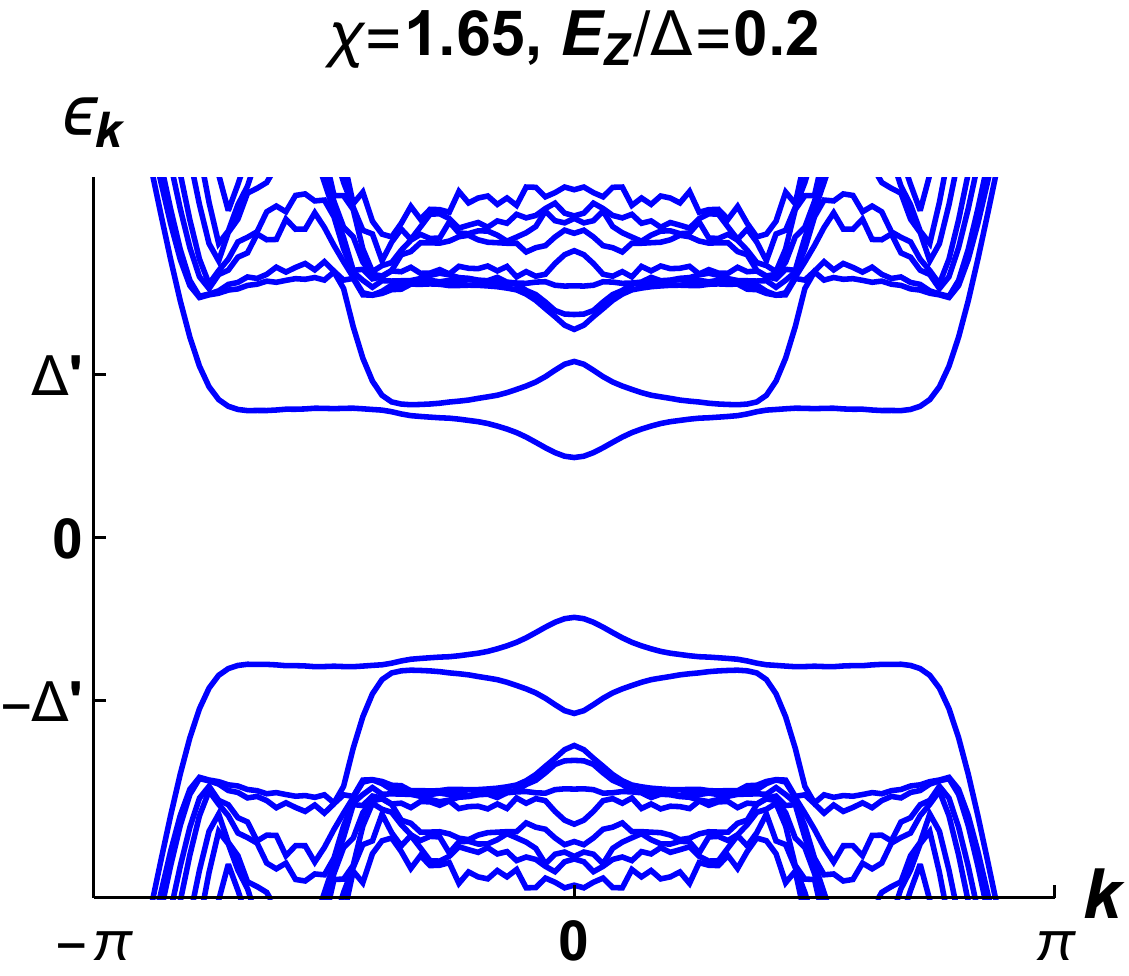}} \quad
	\subfigure[]{
		\includegraphics[height=3cm]{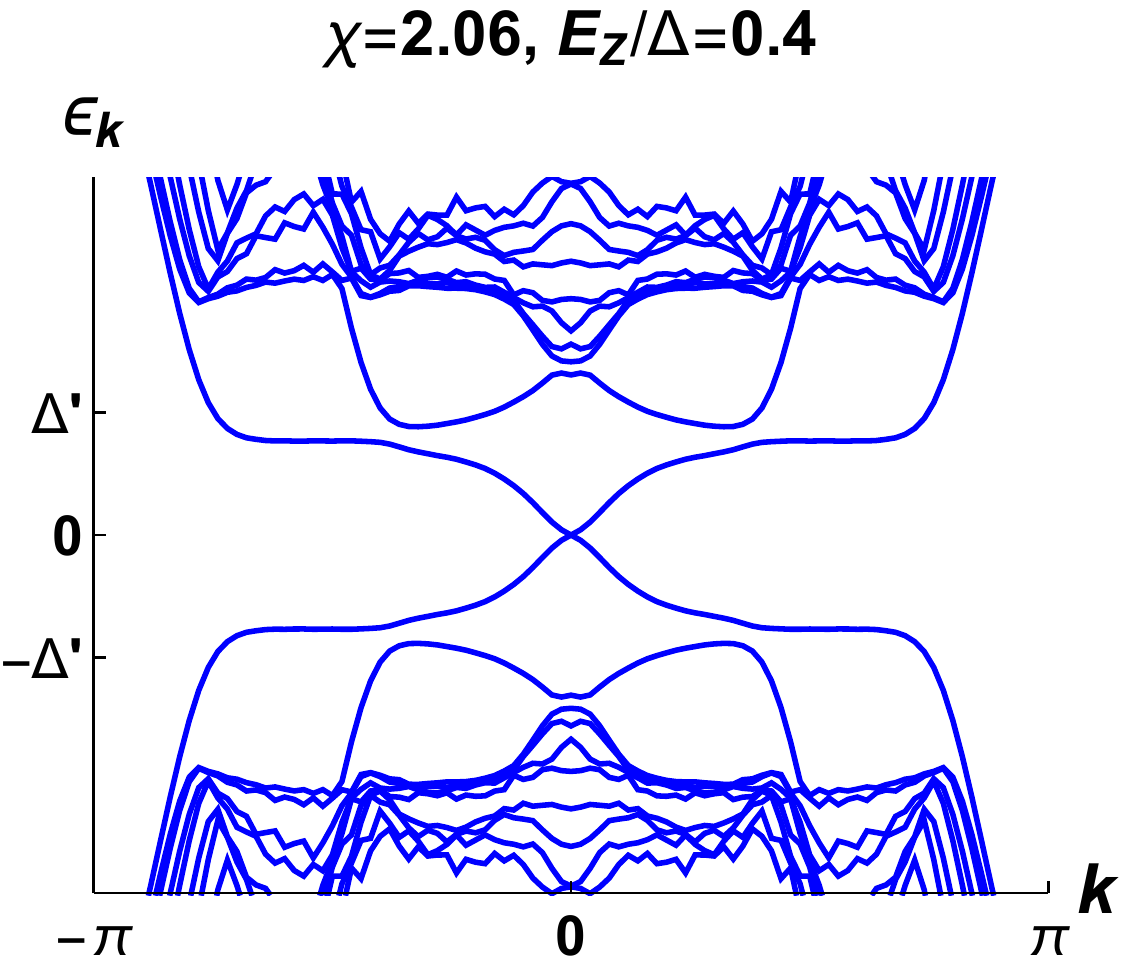}} \quad
	\subfigure[]{
		\includegraphics[height=3cm]{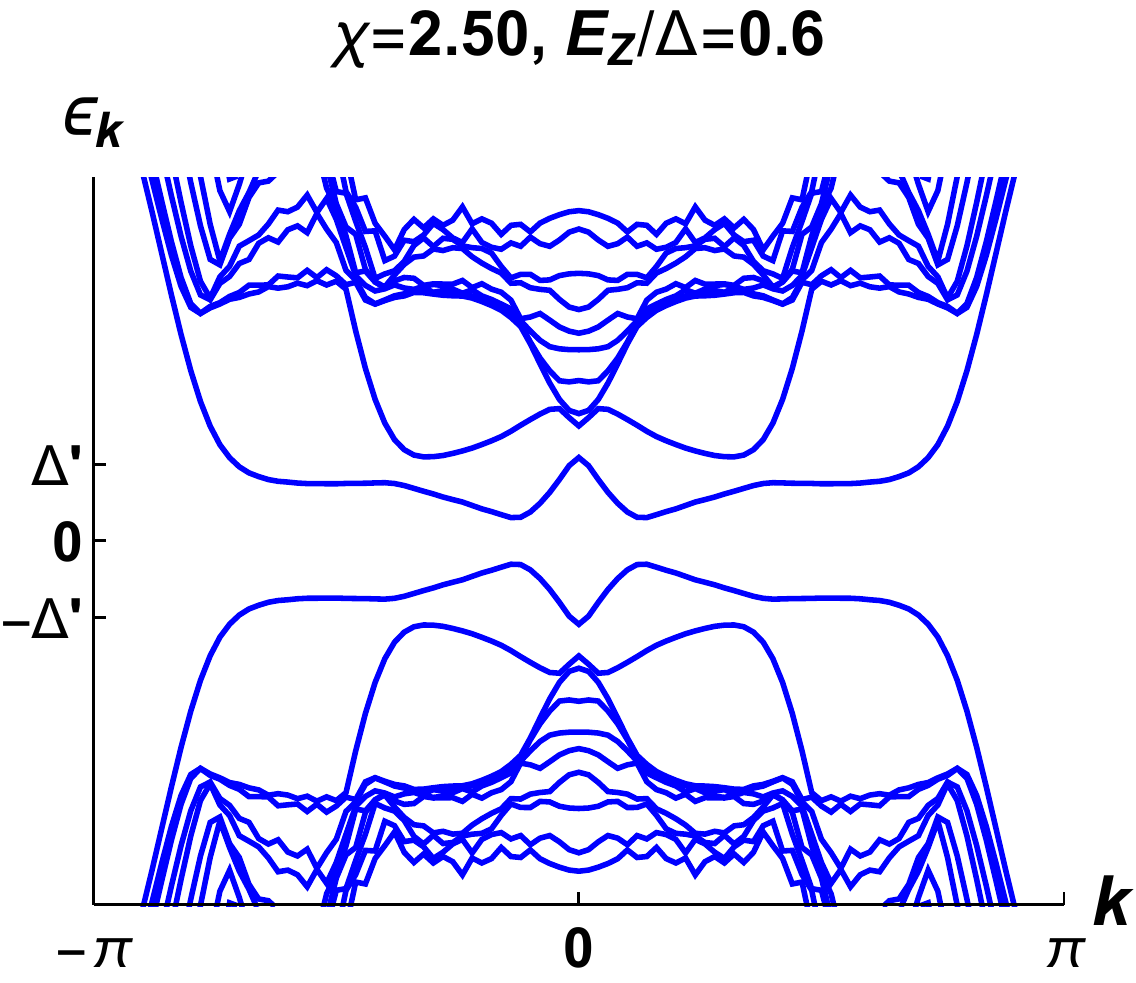}}
	\caption{\label{spectrum}(a) The phase diagram as a function of $(\chi, E_Z/\Delta)$. $C=0$ ($C=1$) is the Chern number indicating the normal SC (the chiral TSC). The solid line is the phase boundary obtained from the lattice Hamiltonian Eq.~(\ref{eq:lattice_hamiltonian}). The parameters are given by $W=25$, $L=2$, $t=t'=1$, $\mu=-3.2$, $\alpha=0.6$, and $\Delta=0.5$. The dashed line is determined by Eq.~(\ref{boundary}) and the dotted line indicates gapless points. The spectra at $k_y=0$ of the Josephson junction at the three red points are shown in (b-d). $\Delta' = \Delta \cos (\chi/2)$ denotes the smallest gap away from the $\Gamma$ point.}
\end{figure*}

\section{Topological gap} \label{sec:gap}

The spectrum at $\bf k=0$ is given by $\epsilon=\pm(\Delta \cos \chi/2 \pm E_Z)$, thus away from the transition point, the gap is still of the same order. 
One does not have to concerned about the gap near the $\Gamma$ point. 
Here, we analyze the topological gap at generic ${\bf k}$ away from the $\Gamma $ point, especially, at $k \gg E_Z/\alpha $, $k = \sqrt{k_x^2 + k_y^2}$. 
Using the scattering theory (see Appendix~\ref{append:gap} for more details), the spectrum at $k \gg E_Z/\alpha $ is determined by
\bea
	\arccos \frac{\epsilon}{\Delta}= \frac{(k_\sigma^e- k_\sigma^h)L}{2} + \frac{\chi}2,
\eea
where $k_\sigma^{e/h}= \sqrt{2m(\mu - E_{k\sigma} \pm \epsilon)}$, $E_{k\sigma}= k^2/2m -\sigma \alpha k$, with $\sigma=\pm 1$. 
In the weak superconducting pairing limit, $\Delta \ll \mu$, the spectrum is
\bea\label{topGap1}
	 \frac{\epsilon}{\Delta} = \cos \Big[ \frac\chi2 + \frac{k_F L}{2\sqrt{1-E_{k\sigma}/\mu}} \frac\Delta\mu \times \frac\epsilon\Delta \Big].
\eea
We can see that the dependence on $k$ is suppressed by $\Delta/\mu$ for $ E_Z/\alpha \ll k \ll k_F$. As $k$ increases such that $E_{k\sigma} \sim \mu$, the prefactor of $\Delta/\mu$ diverges. 
Nevertheless, notice that the bound state exists only if $k_\sigma^{e/h}$ is a real number, i.e., $E_{k\sigma} \le \mu -  \epsilon$. So when $E_{k\sigma} = \mu -  \epsilon$, the spectrum changes to
\bea\label{topGap2}
	 \frac{\epsilon}{\Delta} = \cos \Big[ \frac\chi2 + \frac{k_F L}{2} \sqrt{\frac\Delta\mu} \times \sqrt{\frac\epsilon\Delta} \Big].
\eea
Combining the results of Eqs.~(\ref{topGap1}) and~(\ref{topGap2}), we conclude that the topological gap is $\epsilon \sim \Delta \cos \frac{\chi}2$ up to a small correction of $\Delta/\mu$. 
This gap is also verified in the lattice model introduced in the next section.
Notice that at $\chi = \pi$, the system has a small gap, so the better place to get the chiral topological superconductor is in between $\chi = 0$ and $\pi$, where the Zeeman field is small while the gap is sizable.

\begin{figure*}[htbp]
	\centering
	\subfigure[]{\label{edge1}
		\includegraphics[width=3.7cm]{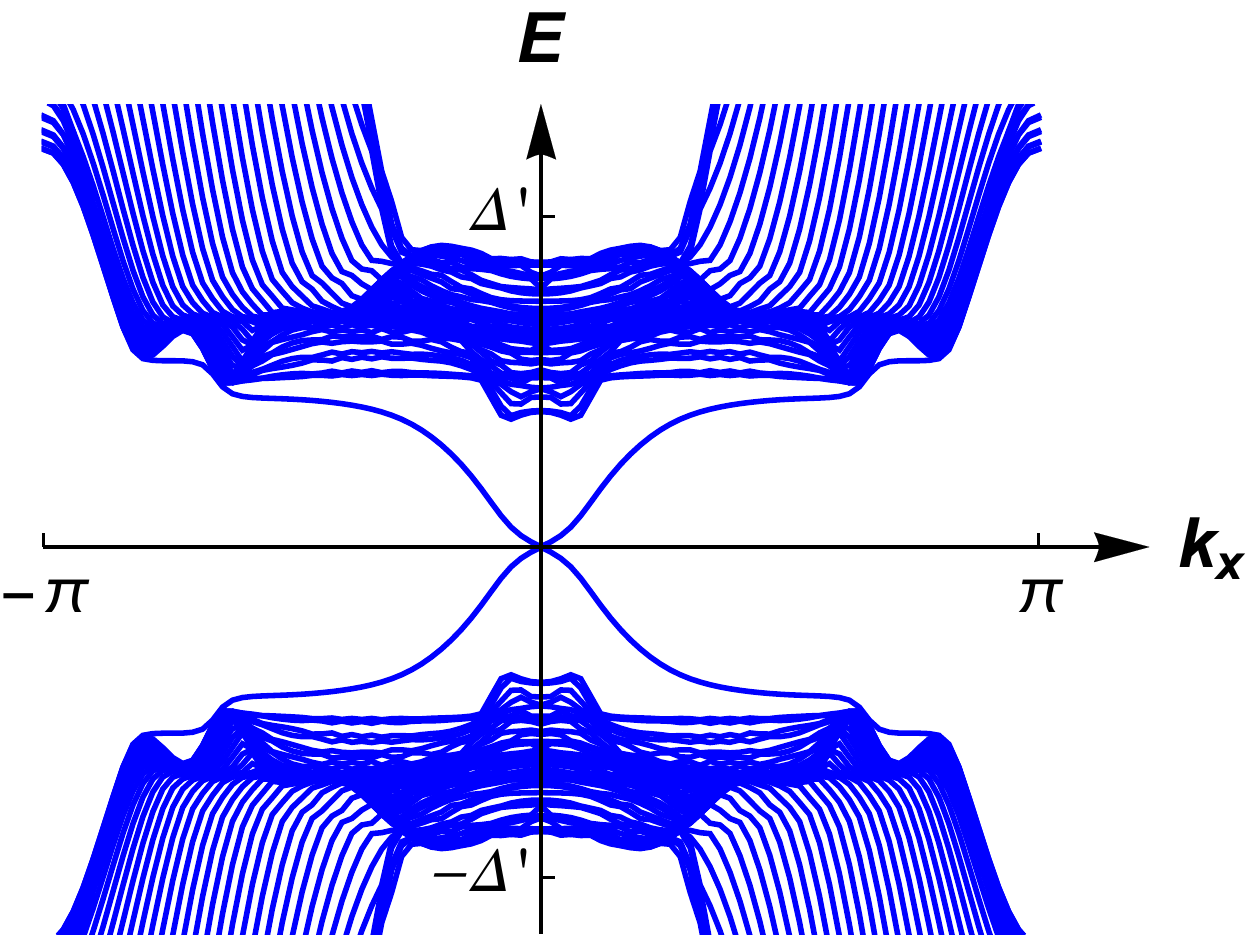}} \quad
	\subfigure[]{\label{edge2}
		\includegraphics[width=3.7cm]{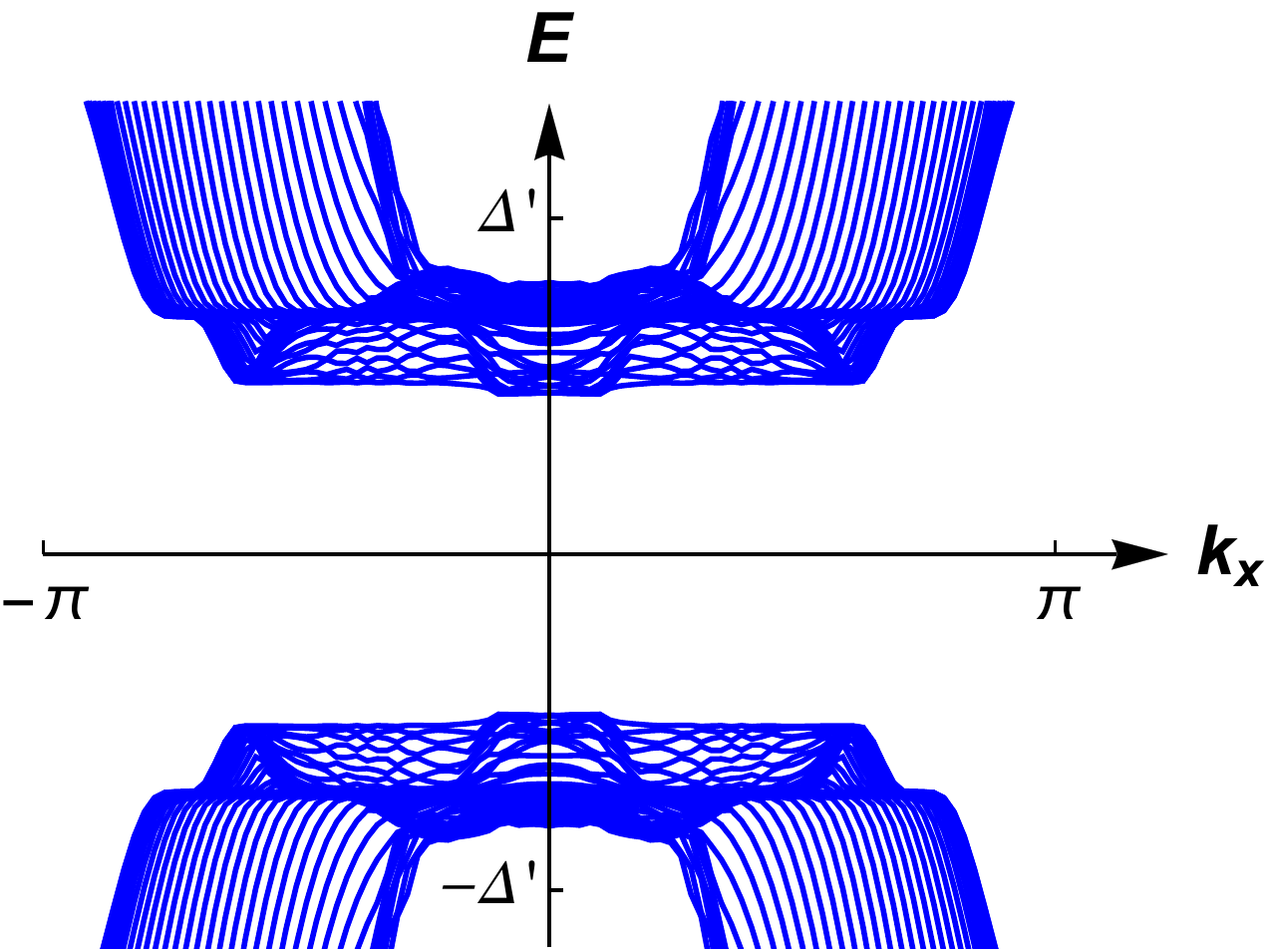}} \quad
	\subfigure[]{\label{vortex}
		\includegraphics[width=3.7cm]{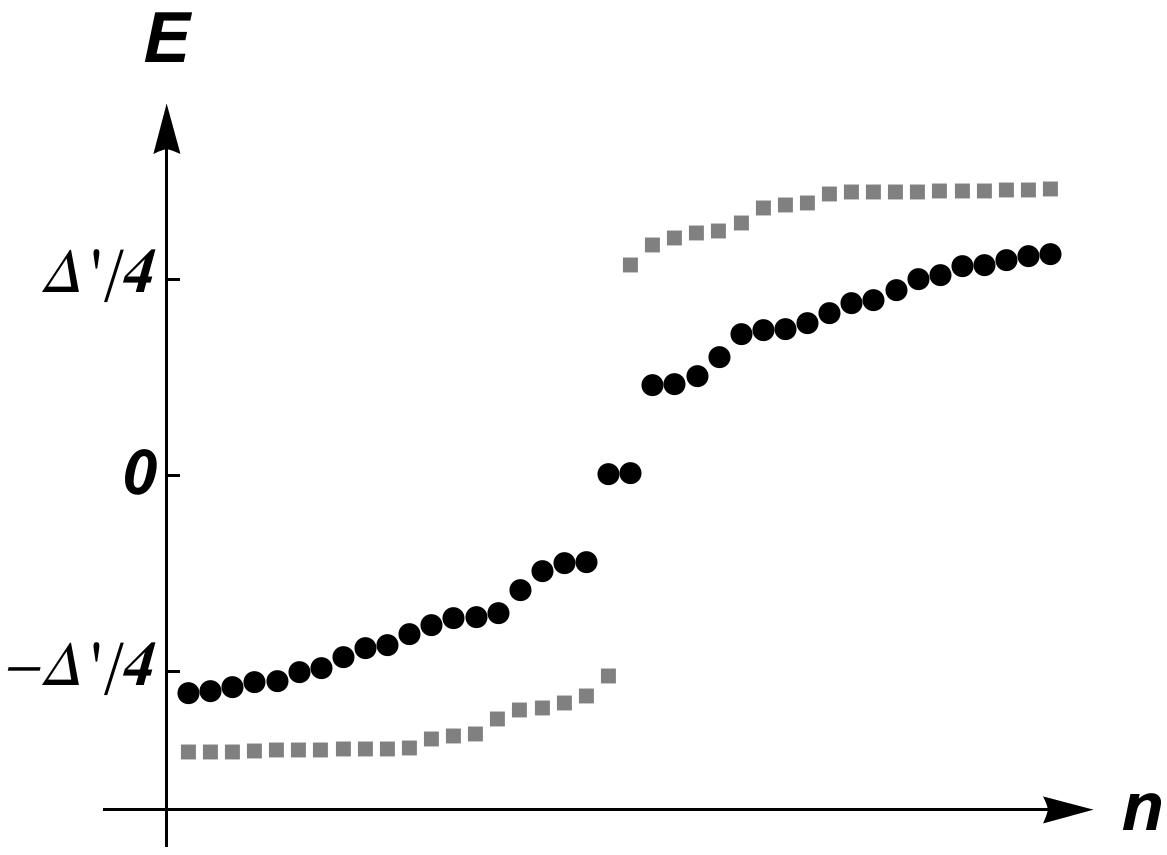}} \quad
	\subfigure[]{\label{density}
		\includegraphics[width=3.7cm]{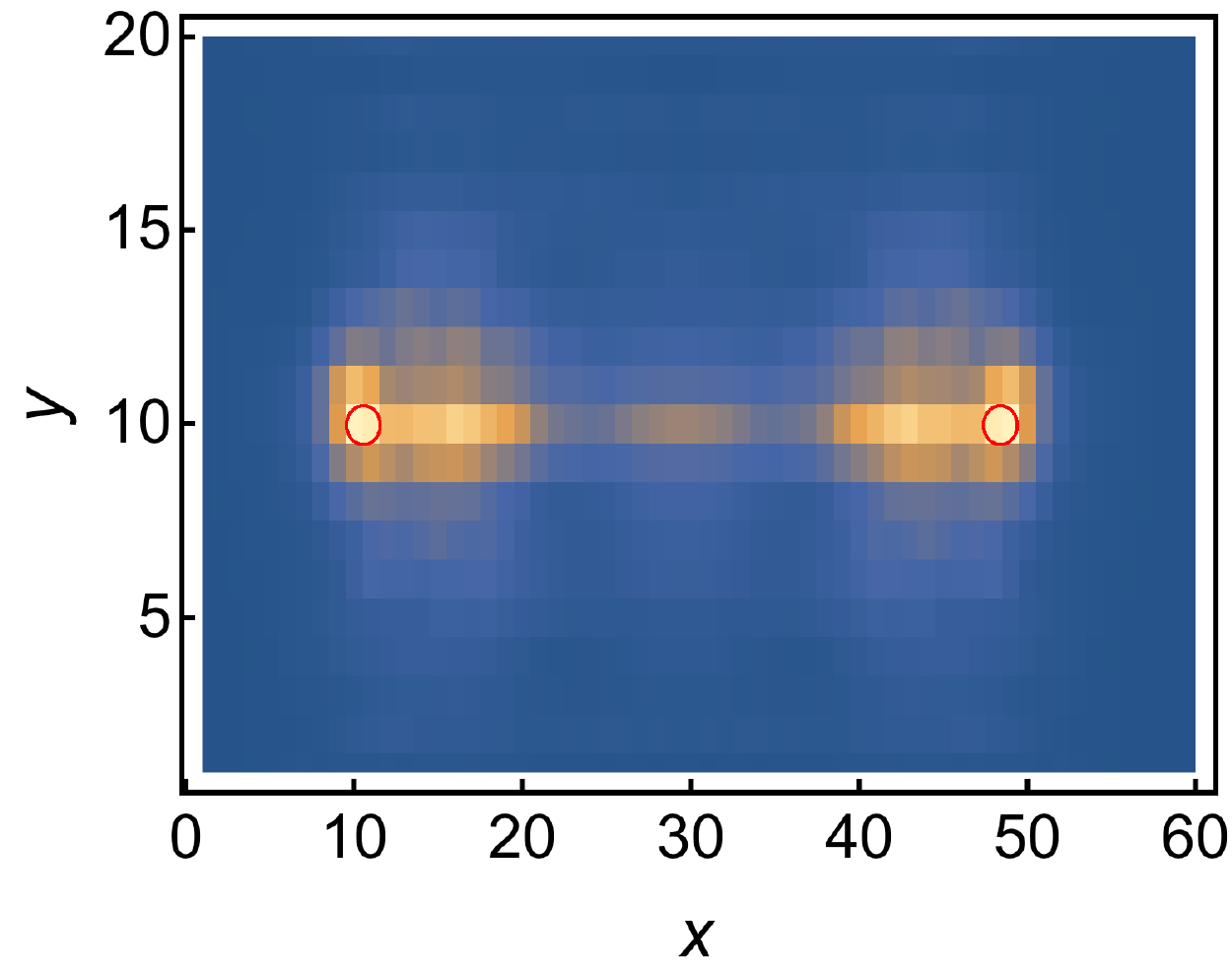}}
	\caption{(a,b) The spectrum of the system in $k_x$ direction (open boundary condition in the $y$ direction) in the topological superconducting phase and the trivial phase, respectively. The parameters are given by $W=5$, $L=2$, $t=t'=1$, $\mu=-3.2$, $\alpha=1$, $\Delta=1$, $\chi=0.6 \pi$. (c) The spectrum of the system with two vortices.  The disk (square) points correspond to the topological (trivial) phase. The parameters are given by $W=6$, $L=2$, $t=t'=1$, $\mu=-3.2$, $\alpha=1$, $\Delta=1$, $\chi=0.6 \pi$. (d) The local density of the ground state wavefunction in the topological phase. Two Majorana zero modes appear in the vortices. The topological (trivial) phase corresponds to $E_Z=0.8 \Delta$ ($E_Z=0.1\Delta$).}
\end{figure*}

\section{Tight binding model} \label{sec:lattice}

To demonstrate the topological phase, we construct a tight binding Hamiltonian to model the setup in Fig.~\ref{josephson}. 
The lattice model is given by $H= H_{0} + H_{SOC} + H_{Z} + H_{SC}$, where
\begin{widetext}
\bea \label{eq:lattice_hamiltonian}
	&& H_0 = - t \sum_{j_z=1}^{W} \sum_{{\bf i}; \hat \mu= \hat x, \hat y} (c_{{\bf i}, j_z}^\dag c_{{\bf i}+\hat \mu, j_z} + H.c. ) - t' \sum_{j_z=1}^{W-1} \sum_{{\bf i}} (c_{{\bf i}, j_z}^\dag c_{{\bf i}, j_z+1} + H.c. )- \mu \sum_{j_z=1}^{W} \sum_{{\bf i}} c_{{\bf i}, j_z}^\dag c_{{\bf i},j_z} , \\
	&& H_{{SOC}} = i \alpha \sum_{j_z=1}^{W} \sum_{{\bf i}; \hat \mu=\hat x,\hat y}  [ (c_{{\bf i}, j_z}^\dag \sigma^x c_{{\bf i}+\hat x, j_z} - c_{{\bf i}, j_z}^\dag \sigma^y c_{{\bf i}+\hat y, j_z}) - H.c. ],  \\
	&& H_{Z} = E_{Z} \sum_{j_z=1}^{W} \sum_{{\bf i}}  c_{{\bf i}, j_z}^\dag \sigma^z c_{{\bf i}, j_z}, \\
	&& H_{SC} =   \sum_{{\bf i}}  \Big[ \sum_{j_z=1}^{[W/2]-[L/2]} \Delta e^{-i\chi/2}  c_{{\bf i}, j_z}  (i \sigma^y) c_{{\bf i}, j_z} + \sum_{j_z=W+1-([W/2]-[L/2])}^{W} \Delta e^{i\chi/2}  c_{{\bf i}, j_z}  (i \sigma^y) c_{{\bf i}, j_z} \Big] + H.c..
\eea
\end{widetext}

Here, $\bf i $ denotes sites in square lattice in the $xy$ plane, and $j_z$ is the layer index along the $z$ direction. 
The number of semiconductor layer is $W$. $t$ ($t'$) is the in-plane (out-of-plane or $z$-direction) nearest-neighbor hopping amplitude. 
$\mu$ is the chemical potential, $\alpha$ is the strength of in-plane spin-orbit coupling, and $\Delta$ ($\chi$) is the amplitude (phase difference) of the proximatized superconducting orders that are nonzero at $1 \le j_z \le [W/2]-[L/2]$ and $ W+1 - ([W/2]-[L/2]) \le j_z \le W$. $[x]$ denote the largest integer that is smaller than $x$.

The phase diagram of the lattice model is shown in Fig.~\ref{phaseDiagram3}. 
The solid line in Fig.~\ref{phaseDiagram3} is the phase boundary obtained from the lattice model, while the dashed line is determined by Eq.~(\ref{boundary}). 
One can see that they match each other quite well. 
The spectrum of the junction at different phases indicated by the red points in Fig.~\ref{phaseDiagram3} are plotted in Figs.~\ref{spectrum}(b-d), where one can see that by tuning the phase difference and/or the Zeeman energy, the energy gap will close at phase boundary and then reopen in the topological superconducting phase.
It is also important to note that the smallest gap away from the $\Gamma$ point is given by $\Delta' = \Delta \cos \frac\chi2$, which is consistent with the result from the scattering theory in the previous section.

To explore the chiral edge state of topological superconducting phase, we use open boundary condition in the $y$ direction, and plot the energy spectrum as a function of $k_x$ as shown in Figs.~\ref{edge1} and~\ref{edge2} for the topological phase and the trivial phase, respectively. 
The two chiral Majorana states along $+y$ and $-y$ edges are clearly shown in Fig.~\ref{edge1} in the topological superconducting phase, although there is no edge state at the normal superconducting phase shown in Fig.~\ref{edge2}.

Moreover, when vortices are created via external magnetic field, the chiral topological superconductor is expected to host Majorana zero mode localized in the vortex core. 
In Fig.~\ref{vortex}, we observe the Majorana zero mode in the topological phase with two vortices, while in the trivial phase there is no Majorana zero mode. 
The local density of ground state wavefunction in the topological phase is plotted in Fig.~\ref{density}, corresponding to the localized Majorana zero mode at the vortex core.

\section{Conclusions} \label{sec:conclusion}

In this paper, we show that a two-dimensional Josephson junction of the superconductor-semiconductor-superconductor sandwich heterostructure can provide a useful setup to realize the chiral topological superconductor with chiral Majorana edge modes. 
The topological superconducting phase appears in a large portion of the phase diagram spanned by the Zeeman field and the superconducting phase difference of the Josephson junction. 
Compared to the previous setup~\cite{Sau2010}, although the device in our proposal might be a little more complicated as the superconductor layers cover both the top and the bottom of the semiconductor, the benefit is that the Zeeman field is reduced within the critical strength and gating is no longer necessary which is considered as the main challenges of previous setup.
For the simplest setup of a Rashba semiconductor proximitized with a superconductor, we can model it by the following BdG Hamiltonian
\bea
	&& H_{BdG} = \left( -t (\cos k_x + \cos k_y) - \mu \right) \tau^z \nn \\
	&& + \alpha ( \sin k_x \sigma^y \tau^z - \sin k_y \sigma^x) + E_Z \sigma^z \tau^z + \Delta \sigma^y \tau^x,
\eea
where $\sigma$ ($\tau$) acts on the spin (Nambu) space. $t$ is the hopping amplitude, $\mu$ is the chemical potential, $\alpha$ is the spin-orbital coupling strength, $E_Z$ denotes the Zeeman energy, and $\Delta$ is the proximitized superconducting gap.
This setup is able to induce a gap closing at the $\Gamma $ point and render a topological transition into the chiral topological superconductivity. 
It is easy to realize the transition occurs at $E_Z = \sqrt{(2t + \mu)^2 + \Delta^2} $. 
So, to realize the topological phase in this setup, the Zeeman energy cannot be less than the superconducting gap, which is typically hard to achieve.
Besides that, to really lower the critical Zeeman energy, one needs to tune the chemical potential.
For our setup, however, as demonstrated in Fig.~\ref{phaseDiagram1}, the topological phase can be obtained by a Zeeman energy that is much smaller the superconducting gap and the chemical potential does not play any essential role.
These advantages facilitate possible experimental reaches. 

We mainly consider the narrow Josephson junction limit, i.e., $\Delta \ll E_T$. 
A wide Josephson junction with $\Delta \gg E_T$ can also achieve the desired topological transition. 
In this case, there appears multiple bound states in the Josephson junction, and the topological phase transition occurs at
\bea
	 \frac{\tilde{\chi}}2 \pm \frac{\pi}2 \frac{E_Z}{E_T} = \left(n+\frac12 \right) \pi.
\eea

One may have concerns that the external magnetic field might not be efficient when the semiconductor is covered from both sides by the superconductors. 
It is useful to consider type-II superconductors in the setup so that the magnetic field can penetrate through the full device by creating vortices. 
And in particular, we have demonstrated by the lattice Hamiltonian that the vortex is not only harmless to the topological phase, but also useful in generating Majorana zero modes, as shown in Fig.~\ref{vortex} and~\ref{density}.
Moreover, instead of using the external magnetic field, it is also possible to consider doping magnetic elements or using the ferromagnetic semiconductors to render the Zeeman energy in the Josephson junction.
The flexibility of achieving the topological phase transition and the mature fabrication technique of quasi-two dimensional semiconductor devices make our proposal a promising setup to realize the chiral topological superconductor. 

Possible experimental signatures include the half-integer longitudinal conductance plateau~\cite{Wang2015}, which is implemented in previous experiments~\cite{He2017, Chang2019}, and zero bias peak in the tunneling conductance of Majorana zero modes trapped in vortices, which we have explicitly shown in Fig.~\ref{density}.

\section{Acknowledgement} 
We thank Hong Yao, Sankar Das Sarma, Ady Stern, Zhongbo Yan, Yingyi Huang, Fuchun Zhang, and Ruirui Du for helpful discussions. S.-K.J. is supported by the Simons Foundation via the It From Qubit Collaboration. S.Y. is supported by the National Natural Science Foundation of China (Grant No. 41030090).

\begin{appendix}

\begin{widetext}

\section{Spectrum in Josephson junction} \label{append:scattering}
Due to the proximity effect, the superconducting (SC) order will penetrate into semiconductor layers. Because the two superconductors have different SC order (most importantly, the different SC phase), there is a domain wall in the semiconductor layers. Assuming the size of the domain wall is $L$ as shown in Fig.~1(b), the SC order in the semiconductor layers is given by
\bea\label{S_SC}
\Delta(z) =  \Delta e^{-i\chi/2} \Theta(-z-L/2) + \Delta e^{i\chi/2} \Theta(z-L/2),
\eea
where $\Delta$ is the magnitude of superconductivity, and $\chi$ denotes the phase difference between two superconductors. $\Theta$ is the step function. Note the difference between the width of semiconductor layers denoted by $W$, and the width of domain wall denoted by $L$, as shown in Fig.~\ref{S_junction}. Because the proximity length of SC order is quite large, and the proximitized SC order in semiconductors opens up a finite energy gap, the amplitude of wavefunctions will concentrate near the domain wall. As a result, one focus on the interface at $\pm L/2$ and take $W/L \rightarrow \infty$.

The BdG Hamiltonian in the semiconductor layers is $H = \Psi^\dag \mathcal H_{BdG} \Psi$, where $\Psi= (\psi_\uparrow, \psi_\downarrow, \psi^\dag_\downarrow, - \psi^\dag_\uparrow)$, and
\bea\label{S_HBdG}
H_{BdG}(k_x, k_y, z) = \Big(\frac{k_x^2 + k_y^2 - \partial_z^2}{2m} -\mu\Big) \tau^z + H_{SOC}(k_x, k_y) \tau^z + (\Delta(z) \tau^+ + H.c.) + E_Z \sigma^z ,
\eea
where $\tau$ ($\sigma$) is the Pauli matrix acting on the Nambu (spin) space and $\tau^\pm = \frac12(\tau^x \pm i \tau^y)$. $E_Z$ denotes the Zeeman energy. Note that the Zeeman field is along $z$ direction. $H_{SOC}$ is the Dresselhaus spin-orbit coupling which will be discussed in the next Sector.

Owing to the translational symmetry in $xy$-plane, $k_x$ and $k_y$ are good quantum numbers. Let's first focus on $k_x = k_y = 0$ subspace in this Sector to determine the phase diagram, and then in later Sector on generic $k_x, k_y$ to explore the topological gap. The Hamiltonian in the $k_x = k_y = 0$ subspace reduces to
\bea
H_{BdG}(0,0, z) = \Big(-\frac{\partial_z^2}{2m} -\mu\Big) \tau^z + (\Delta(z) \tau^+ + H.c.) + E_Z \sigma^z.
\eea
For the normal state, i.e., $|z|< L/2$, the eigenstates with eigen-energy $\epsilon$ are
\bea\label{normalState}
\psi_{e\sigma}^{\pm}(z) = \frac1{\sqrt{k_{\sigma}^e}} \left( \ba{cccc} 1 \\ 0 \ea \right) \otimes |\sigma\rangle e^{ \pm i k_{\sigma}^e z},
\quad
\psi_{h\sigma}^{\pm}(z) = \frac1{\sqrt{k_{\sigma}^h}} \left( \ba{cccc} 0 \\ 1 \ea \right) \otimes |\sigma\rangle e^{ \pm i k_{\sigma}^h z},
\eea
where $\sigma=\pm 1$, $  |+1 \rangle = (1,0)^T$ and $|-1 \rangle = (0,1)^T$. The wavevector is $k_\sigma^{e/h}= \sqrt{2m(\mu \pm (\epsilon - \sigma E_Z))}$. For the proximitized SC state at $|z|>L/2$, the eigenstates with eigen-energy $\epsilon$ is give by
\bea\label{SCState}
\phi_{e\sigma}^{\pm}(z) = \frac1{\sqrt{2q_{\sigma}^e}[(\epsilon-\sigma E_Z)^2/\Delta^2-1]^{1/4}}  \left( \ba{cccc} e^{i\eta_\sigma^e/2} \\ e^{-i\eta_\sigma^e/2} \ea \right) \otimes |\sigma\rangle e^{ \pm i q_{\sigma}^ez},
\eea
and $\phi_{h\sigma}^{\pm}$ is given by replacing the index $e$ to $h$ in Eq.~(\ref{SCState}). Here $\eta_\sigma^{e/h} = \frac{\chi}2 \pm \eta_\sigma$, and $\eta_\sigma = \arccos \frac{\epsilon- \sigma E_Z}{\Delta}$. Notice that for $t>1$, $\arccos t = -i \ln(t+ \sqrt{t^2-1})$. The wavevector is $q_\sigma^{e/h}= [2m(\mu \pm \sqrt{(\epsilon - \sigma E_Z)^2 - \Delta^2})]^{1/2}$, and $\Re q_\sigma^{e/h} >0$, $\Im q_\sigma^{e} >0$, $\Im q_\sigma^{h}<0$. The wavefunctions are normalized to have same current.

We are ready to evaluate the scattering matrix. To simplify the problem, it is easy to observe that the spin is conserved during the scattering. So let's focus on $\sigma=1$ and neglect the index $\sigma$. By matching the wavefunction Eq.~(\ref{normalState}) and Eq.~(\ref{SCState}) at the superconducting-normal state interface at $z= \pm L/2$, the scattering matrix is $S_{\pm L/2}= e^{\pm i \chi/4 \tau^z} S e^{\mp i \chi/4 \tau^z}$, where
\bea
S = e^{-i\phi_S} \left( \ba{cccc} r e^{i\phi_N} & \sqrt{1-r^2} \\
\sqrt{1-r^2} & -r e^{-i\phi_N} \ea \right),
\eea
in which $r= r_n/r_d$ is normal reflection amplitude. 
$r_n$ and $r_d$ are positive numbers, and
\bea
r_d e^{i \phi_S} =-e^{-i \eta}(k^h-q^e)(k^e-q^h)+e^{i \eta}(k^e+q^e)(k^h+q^h),\\
r_n e^{i \phi_N} = -e^{-i \eta}(k^h-q^e)(k^e+q^h)+e^{i \eta}(k^e-q^e)(k^h+q^h).
\eea
These functions determine the two phases $\phi_S$ and $\phi_N$ in Eq.~(\ref{eq:scattering_matrix}) in the main text.

The transmission amplitude is simply
\bea
T = \left( \ba{cccc} e^{i k^e L} & 0 \\
0 & e^{-i k^h L} \ea \right).
\eea
Using the scattering and transmission matrix, the spectrum of the bound state is determined by $\det(1- S_{-L/2}TS_{+L/2}T)=0$, i.e.,
\bea
\cos[(k^e-k^h)L-2\phi_S] = (1-r^2) \cos \chi + r^2 \cos[(k^e + k^h)L + 2\phi_N].
\eea

To proceed, we assume weak pairing $E_Z \lesssim \Delta \ll \mu$, then we have $r= \Delta/(2\mu) + \mathcal{O}(\Delta^2/\mu^2)$, $\phi_S = \eta$, and $\phi_N=0$. The zero mode is given by
\bea
\arccos \frac{E_Z}{\Delta}= \frac{\tilde{\chi}}2 + \frac{\pi}2 \frac{E_Z}{E_T} + n\pi,
\eea
where $\tilde \chi $ is a function of $\chi$:
\bea\label{S_chi}
\tilde\chi(\chi) = \arccos \Big(\big[1- \frac{\Delta^2}{(2\mu)^2} \big] \cos \chi + \frac{\Delta^2}{(2\mu)^2} \cos 2 k_F L \Big),
\eea
and $E_T \equiv (\pi/2) v_F/L$ is the Thouless energy of the junction. Here $k_F = \sqrt{2m \mu}$ and $v_F = \sqrt{2\mu/m}$. Note the domain wall width is of order of lattice constant, as a result, the Thouless energy is of order of chemical potential, $E_T \sim \mu$, which is the largest energy scale in the question.

\begin{figure}
	\includegraphics[width=5cm]{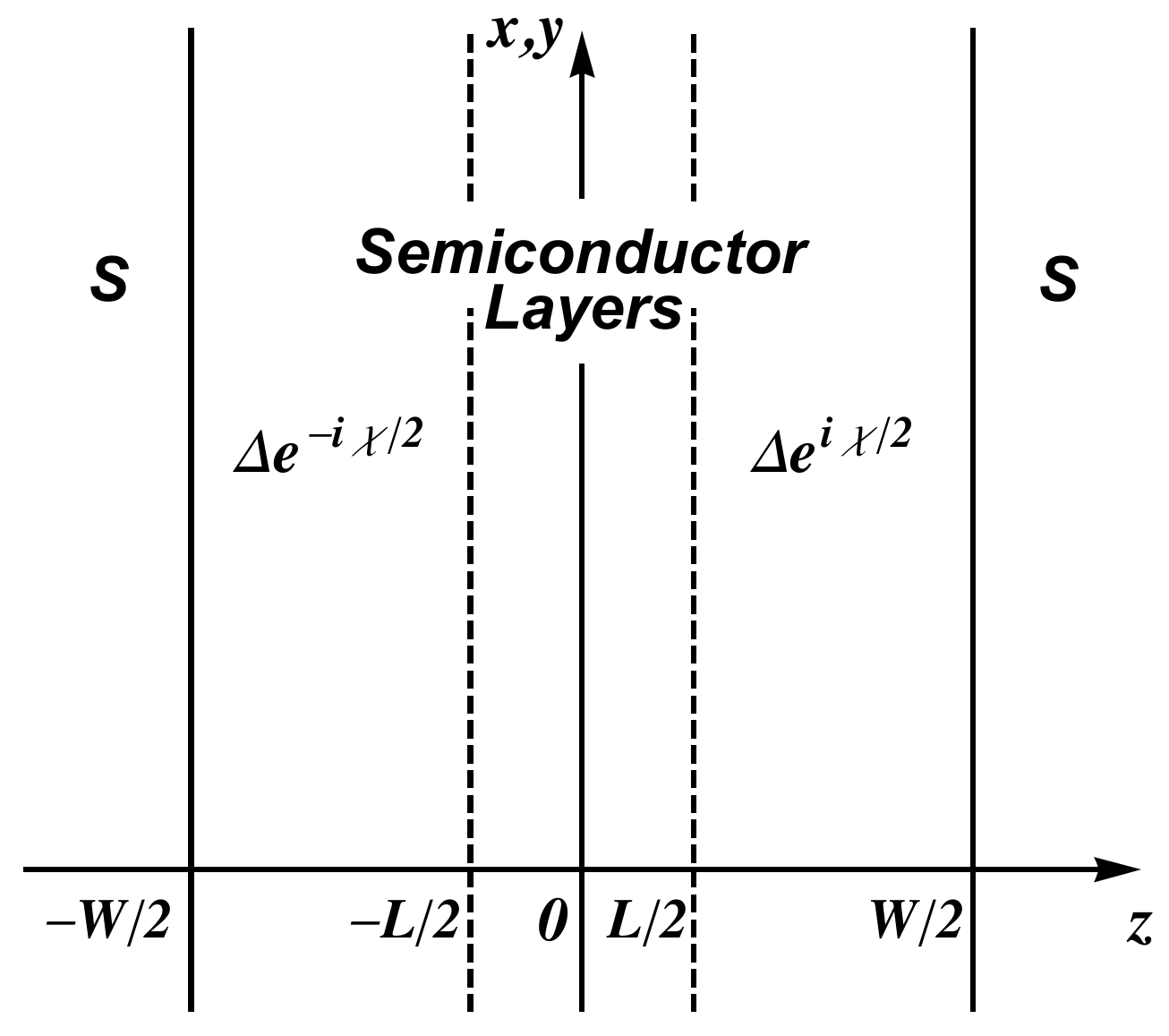}
	\caption{\label{S_junction}Schematic plot of Josephson junction.}
\end{figure}

As we will show in the next Section, the appearance of zero mode implies a gap closing that separates trivial and topological superconducting phases. At the zeroth order of $\Delta/\mu$, there is no normal reflection and $\tilde\chi = {\chi}$, the phase diagram is simply given by $\frac{E_Z}{\Delta} = |\cos \frac{\chi}2 |$, as shown by the dashed line in Fig.~2. It is interesting to see that at $\chi=\pi$, an infinitesimal magnetic field can turn the system into a chiral topological superconducting phase.

Including correction from the normal reflection to the first order of $\Delta/\mu$, the relation between $\chi$ and $\tilde\chi$ is given in Eq.~(\ref{S_chi}) and the phase diagram is corrected to the solid line in Fig.~2. At $\chi=0$ and $\chi=\pi$, the correction is in the linear order of $\Delta/\mu$, i.e., $\tilde\chi(\chi=0) = \Delta/\mu \sin k_F L$, and $\tilde{\chi}(\chi=\pi) = \pi - \Delta/\mu \cos k_F L$, respectively. While at other generic values of $\chi$, the correction is in the quadratic order of $\Delta/\mu$, i.e., $\tilde\chi(\chi) = \chi + (\Delta/\mu)^2(\cos\chi- \cos2k_F L)/\sin \chi$. It is useful to note that when the normal reflection is considered, the critical Zeeman energy at $\chi=\pi$ is given by $\Delta^2/\mu$, which is a still a small number at weak pairing.

\section{Spin-orbit coupling and topological phase transition} \label{append:transition}

It is well known that spin-orbit coupling is crucial in the realization of topological phases. Direct semiconductors with an inversion-asymmetric zinc blende structure (point group $T_d$), such as GaAs, InSb, CdTe etc, have considerable size of spin-orbit coupling and are common in making quantum structures. Moreover, many of these materials have a similar band structure with the smallest gap at $\Gamma$ point. To give a concrete example of spin-orbit couplings, we consider a heterostructure consisting of a few layers of such semiconductors grown in (001) direction. Because of the phase difference in the Josephson junction, the heterostructure breaks inversion symmetry. However, such a structure inversion asymmetry cannot induce Rashba spin-orbit coupling, since the system is invariant under the composite $S_4$ and time reversal operations. As a result, the symmetry allowed spin-orbit coupling is the Dresselhouse term
\bea
H_{SOC} = \alpha (k_x \sigma^x - k_y \sigma^y),
\eea
where $\alpha$ is the strength of spin-orbit coupling.

In the following, we will show that the gap closing in the pervious Section is described by two-dimensional Majorana cone. To simplify the question, we set the width of domain wall $L$ to be zero, and neglect the normal reflection. In the $k_x= k_y =0 $ subspace of the semiconductor (without proximitized SC order), the dispersion along $k_z$ direction is shown in Fig.~\ref{dirac}. We make a continuous model near the Fermi point, i.e., $\psi_{1,2}$, for right and left movers. Because the proximitized SC order couples the Kramers pair, we first analyze the Kramers pair $\psi_{1,\uparrow}$ and $\psi_{2,\downarrow}$ with kinetic term given by one-dimensional Dirac Hamiltonian $\mathcal H_0 = -i v_F \partial_z \sigma^z$. 

Owing to the proximitized SC order, the BdG Hamiltonian in Nambu space $\Psi = (\psi_{1,\uparrow, z}, \psi_{2,\downarrow, z}, \psi_{1,\uparrow, z}^\dag, \psi_{2,\downarrow, z}^\dag)^T$ is given by (note the basis difference in Eq.~(\ref{S_HBdG})), $\mathcal{H}_{BdG}(k_x, k_y, z) = H_0(z) + H_1(k_x ,k_y)$, where
\bea
&& H_0(z) = -i v_F \partial_z \sigma^z - \Delta'(z) \sigma^y \tau^y - \Delta''(z) \sigma^y \tau^x  + E_Z \sigma^z \tau^z, \label{S_H0} \\
&& H_1(k_x, k_y) = \frac{k_x^2 + k_y^2}{2m} \tau^z + \alpha(k_x \sigma^x - k_y \sigma^y \tau^z),
\eea
and $\Delta'(z)= \Delta \cos(\chi/2)$ and $\Delta''(z) = \Delta \sin (\chi/2) \text{sgn}(z)$ refer to the real and imaginary part of SC order given in Eq.~(\ref{S_SC}). Here, we have separated the Hamiltonian into two parts, $H_0$ and $H_1$, which will be treated as unperturbed part and perturbation, respectively.

\begin{figure}
	\centering
	\includegraphics[width=4cm]{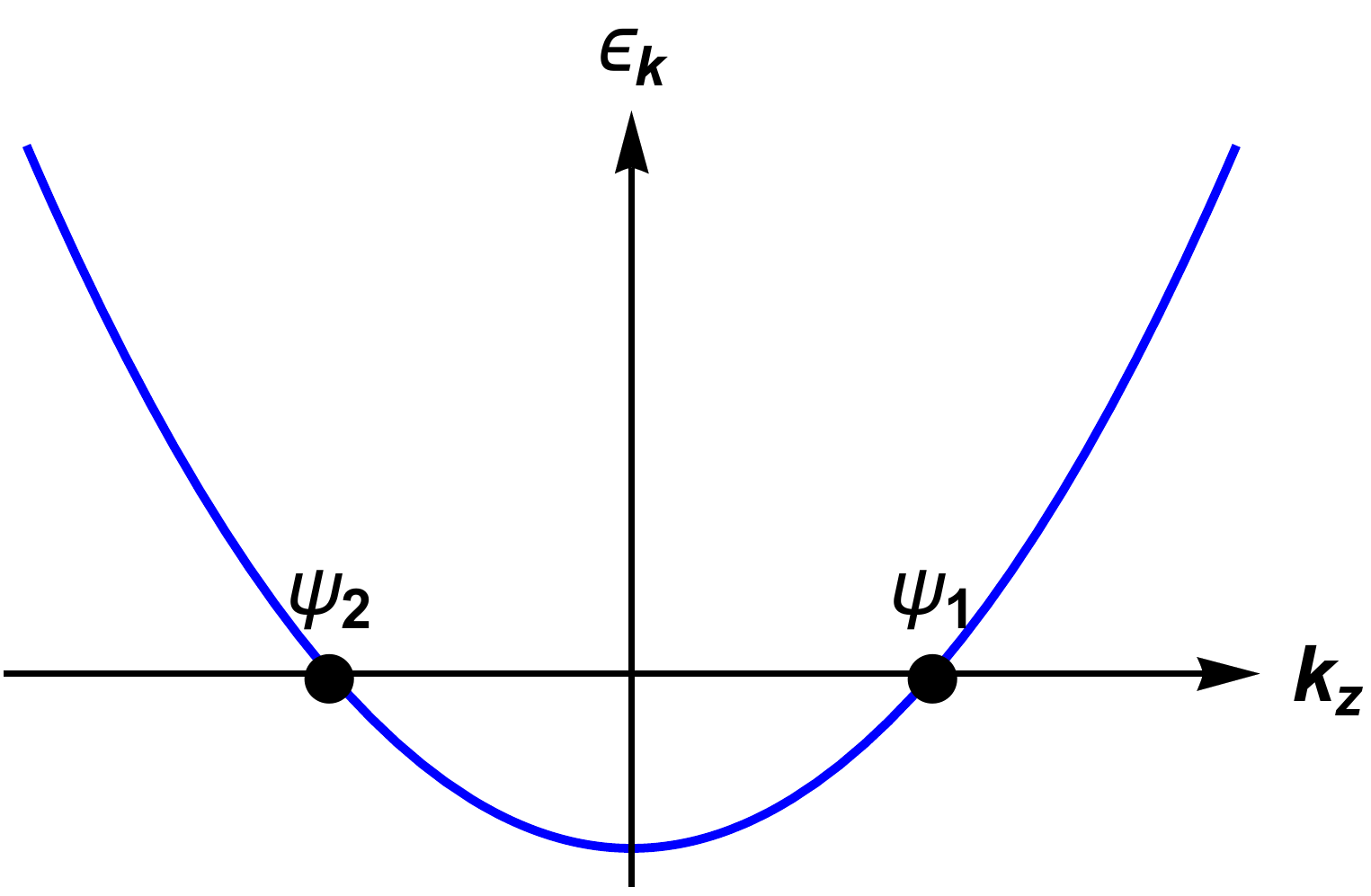}
	\caption{\label{dirac} The dispersion of $k_x= k_y=0$ subspace along $k_z$ direction. Because in $k_x= k_y=0$ subspace the SOC vanishes, the band is doubly degenerate. We denote left/right mover by $\psi_{1/2}$, respectively.}
\end{figure}

In $H_0$, because of the domain wall formed in $\Delta''(z)$, similar to domain wall in one-dimensional Dirac Hamiltonian, we get two related bound states that are related by particle-hole transformation,
\bea
\psi_+(z) = f(z)  (1, 0, 0, -1)^T, \quad \psi_-(z) = f(z)  (0, 1, -1, 0)^T,
\eea
where $f(z) = (\mathcal N)^{-1/2}\exp[-\int_0^z dz' \frac{\Delta''(z')}{v_F}]$, and $\mathcal N$ is the normalization factor. The eigen-energies of these bound states are $ E_{\pm}= \pm (E_Z - \Delta \cos \chi/2)$, and the zero-mode appears at $E_Z/\Delta = \cos \chi/2$. 

Now we consider the perturbation $H_1$ to the bound state $\psi_\pm$. Using first-order perturbation theory, it is straightforward to get the effective Hamiltonian in the bound states subspace $(\psi_+, \psi_-)^T$,
\bea
\mathcal H_\text{eff} = \left(  \ba{cccc} \delta m & -\alpha (k_x+ i k_y) \\
-\alpha (k_x - i k_y) & -\delta m \ea \right),
\eea
where the mass is $\delta m = E_Z - \Delta \cos \chi/2 $. At critical point $\delta m=0$, the Hamiltonian describes a Majorana cone, indicating a topological transition. Namely, the gap closing realizes a transition from a trivial to a topological superconductor.

Similar analysis can be applied to the other Kramers pair $\psi_{1,\downarrow}$ and $\psi_{2,\uparrow}$ in Fig.~\ref{dirac}, and the transition appears at $E_Z/\Delta = -\cos \chi/2$. Combined with above results, the phase boundary is $E_Z/\Delta = |\cos \chi/2|$, consistent with the results from the previous Section.

\section{Topological gap} \label{append:gap}

In previous sections, we have analyzed the spectrum at $k_x=k_y=0$. Away from the transition point, the gap at $k_x=k_y=0$ is in the order of $\Delta$. Here, we analyze the topological gap at generic $k_x, k_y$, especially, when $k_x, k_y \gg E_Z/\alpha $. Since we are interested in $\alpha k \gg E_Z$, $k = \sqrt{k_x^2 + k_y^2}$, we neglect the Zeeman energy, and the Hamiltonian is
\bea
H_{BdG}(E_Z=0) = \Big(\frac{k_x^2 + k_y^2 - \partial_z^2}{2m} -\mu + H_{SOC}(k_x,k_y) \Big) \tau^z + (\Delta(z) \tau^+ + H.c.).
\eea
where the proximitized SC order is again given by Eq.~(\ref{S_SC}). In the following, we will use the same notation in Section A. For the normal state in $|z|< L/2$, the eigenstates with eigen-energy $\epsilon$ are given by
\bea\label{normalState2}
\psi_{e\sigma}^{\pm}(z) = \frac1{\sqrt{k_{\sigma}^e}} \left( \ba{cccc} 1 \\ 0 \ea \right) \otimes |\sigma\rangle e^{ \pm i k_{\sigma}^e z},
\quad
\psi_{h\sigma}^{\pm}(z) = \frac1{\sqrt{k_{\sigma}^h}} \left( \ba{cccc} 0 \\ 1 \ea \right) \otimes |\sigma\rangle e^{ \pm i k_{\sigma}^h z},
\eea
where $\sigma=\pm 1$, $|+1\rangle = 1/\sqrt2 (e^{i \phi_k/2}, e^{-i \phi_k/2})^T$ and $|-1 \rangle = 1/\sqrt2 (e^{i \phi_k/2}, -e^{-i \phi_k/2})^T$, $\phi_k = \arctan k_y/k_x$. The wavevector is $k_\sigma^{e/h}= \sqrt{2m(\mu - E_{k\sigma} \pm \epsilon)}$, where $E_{k\sigma}= k^2/2m -\sigma \alpha k$. For the proximitized SC state in $|z|< -L/2$, the eigenstates with eigen-energy $\epsilon$ is give by
\bea\label{SCState2}
\phi_{e\sigma}^{\pm}(z) = \frac1{\sqrt{2q_{\sigma}^e}[\epsilon^2/\Delta^2-1]^{1/4}}  \left( \ba{cccc} e^{i\eta^e/2} \\ e^{-i\eta^e/2} \ea \right) \otimes |\sigma\rangle e^{ \pm i q_{\sigma}^ez},
\eea
and $\phi_{h\sigma}^{\pm}$ is given by replacing the index $e$ to $h$ in Eq.~(\ref{SCState2}). Here $\eta^{e/h} = \frac{\chi}2 \pm \eta$, and $\eta = \arccos \frac{\epsilon}{\Delta}$. The wavevector is $q_\sigma^{e/h}= [2m(\mu - E_{k\sigma} \pm \sqrt{\epsilon^2 - \Delta^2})]^{1/2}$, and $\Re q_\sigma^{e/h} >0$, $\Im q_\sigma^{e} >0$, $\Im q_\sigma^{h}<0$.

It is easy to observe that the spin is conserved during the scattering. By matching the wavefunction Eq.~(\ref{normalState2}) and Eq.~(\ref{SCState2}) at the superconducting-normal state interface $z= \pm L/2$, we can obtain the scattering matrix, and then get the bound state spectrum. The spectrum is determined by
\bea
\arccos \frac{\epsilon}{\Delta}= \frac{(k_\sigma^e- k_\sigma^h)L}{2} + \frac{\chi}2.
\eea
Here we have neglected the normal reflection which is a small correction in the weak pairing. To proceed, we again assume the weak pairing $\Delta \ll \mu$, and get
\bea
\frac{\epsilon}{\Delta} = \cos \Big[ \frac\chi2 + \frac{k_F L}{2\sqrt{1-E_{k\sigma}/\mu}} \frac\Delta\mu \times \frac\epsilon\Delta \Big].
\eea
The dependence on $k$ is suppressed by $\Delta/\mu$. Thus, to the lowest order, the gap is given by $\epsilon = \Delta \cos \chi/2$.

However, as $k$ increases such that $E_{k\sigma} \sim \mu$, above approximation may break down. Nevertheless, notice that the bound state exists when $k_\sigma^{e/h}$ is a real number, i.e., $E_{k\sigma} \le \mu -  \epsilon$. At $E_{k\sigma} = \mu -  \epsilon$, we have
\bea
\frac{\epsilon}{\Delta} = \cos \Big[ \frac\chi2 + \frac{k_F L}{2} \sqrt{\frac\Delta\mu} \times \sqrt{\frac\epsilon\Delta} \Big],
\eea
where the suppression is sublinear in $\Delta/\mu$. At generic momentum, the topological gap is $\epsilon = \Delta \cos \frac{\chi}2$ up to a small correction of $\Delta/\mu$.

\end{widetext}

\end{appendix}

\end{document}